\documentclass[%
 aip,
 amsmath,amssymb,
 reprint,%
]{revtex4-1}

\usepackage{graphicx}
\usepackage{dcolumn}
\usepackage{bm}

\usepackage[utf8]{inputenc}
\usepackage[T1]{fontenc}
\usepackage{mathptmx,xcolor}

\renewcommand{\[}{\begin{equation}}
\renewcommand{\]}{\end{equation}}
\def\bea{\begin{eqnarray}}
\def\eea{\end{eqnarray}}
\def\nn{\nonumber\\}
\newcommand{\equ}[1]{Eq.~(\ref{#1})}
\newcommand{\eqs}[2]{Eqs.~(\ref{#1}) and (\ref{#2})}
\newcommand{\kk}{\mbox{\boldmath$\kappa$}}
\def\v{{\bf v}}
\def\p{{\bf p}}

\def\r{{\bf r}}
\def\R{{\bf R}}
\def\k{{\bf k}}
\def\E{{\bf E}}

\def\P{{\bf P}}

\newcommand{\ei}[1]{{\rm e}^{i #1}}
\newcommand{\emi}[1]{{\rm e}^{-i #1}}
\renewcommand{\r}{{\bf r}}

\def\ket#1{\vert#1\rangle}
\def\ev#1{\langle#1\rangle}
\def\me#1#2#3{\langle#1| \, #2 \, |#3\rangle}

\begin{document}

\preprint{AIP/123-QED}

\title{Perspective: \\ From the dipole of a crystallite to the polarization of a crystal}

\author{Raffaele Resta}
 \email{resta@iom.cnr.it}
\affiliation{ 
Istituto Officina dei Materiali IOM-CNR, Strada Costiera 11, 34151 Trieste, Italy
}%

 \homepage{http://www-dft.ts.infn.it/\~{}resta/.}
\affiliation{Donostia International Physics Center, 20018 San Sebasti{\'a}n, Spain
}%

\date{\today}

\begin{abstract}
The quantum-mechanical expression for the polarization of a crystalline solid does not bear any resemblance to the (trivial) expression for the dipole of a bounded crystallite; and in fact it has been proved via a conceptually different path. Here I show how to alternatively define the dipole of a bounded sample in a somewhat unconventional way; from such formula, the crystalline polarization formula---as routinely implemented in electronic-structure codes---follows almost seamlessly. 
\end{abstract}

\maketitle


\section{Introduction}

The dipole of a bounded and charge-neutral sample is a very trivial quantity; the macroscopic polarization of a crystalline solid, instead, has been a challenging problem for many years. In quantum mechanics the dipole of a bounded sample is the expectation value of the position operator $\r$. The drawback is that solid state physics requires Born-von-K\`arm\`an periodic boundary conditions (PBCs),\cite{Kittel} which define the Hilbert space where Schr\"odinger equation is solved. Unfortunately the multiplicative operator $\r$ is {\it not} a legitimate operator in the PBC Hilbert space: it maps a state vector within the space into an entity which does not belong to the same space.

The ultimate solution of this long-standing problem was arrived at along the 1990s;\cite{rap73,King93,Vanderbilt93,Ortiz94,rap100} by now, polarization theory is a mature topic.\cite{Spaldin12,Vanderbilt} The historical development of the theory passed through abandoning the concept of polarization ``itself'', addressing instead a polarization {\it difference}, which could be expressed as a time-integrated adiabatic current.\cite{rap73,King93} Only afterwards it was realized \cite{Vanderbilt93} that even polarization itself can be defined, although by means of a change of paradigm: bulk polarization is not a vector (as theretofore assumed), it is a lattice. A counterintuitive corollary is that the polarization $\P$ of an inversion-symmetric crystal is not necessarily zero. Since inversion symmetry requires $\P = -\P$, the lattice must be symmetric: this may happen even if $\P=0$ does not belong to the lattice. For a macroscopic bounded crystallite, the lattice ambiguity is fixed only after the sample termination is chosen.\cite{Vanderbilt,rap136}

What is disturbing is that the two definitions of essentially the same observable---dipole of a crystallite vs. polarization of a crystal---do not bear any formal resemblance. A basic tenet of statistical mechanics and condensed matter physics requires instead that crystalline polarization can also be expressed as the large-sample limit of the dipole of a bounded crystallite over its volume. In this work I am going to bridge this conceptual gap; it will be shown that when the dipole of a bounded sample is alternatively expressed in an unconventional way, with no reference to the $\r$ operator, the crystalline expression follows somewhat naturally.

The paper is organized as follows. In Sec. \ref{sec:bounded} I show how to recast the dipole of a bounded sample in an alternative way, where the $\r$ operator no longer appears: \equ{conne} below. In Sec. \ref{sec:crystal} I show that the same expression can be carried over to a PBC framework, after some logical adaptation, in insulators only. Sec. \ref{sec:multi} shows that in the three-dimensional case one needs to exploit crystalline symmetry in order to make $\P$ a uniquely defined multivalued observable. The mean-field expression---as implemented in Hartree-Fock and density-functional codes---is presented in  Sec. \ref{sec:single} as a special case of the general theory.  Finally, in Sec. \ref{sec:conclusions} I draw some conclusions.

\section{Bounded crystallite} \label{sec:bounded}

We assume that $N$ electrons are confined in a macroscopic sample of volume $\cal V$, together with a neutralizing background of point-like classical nuclei. Let $\ket{\tilde\Psi_0}$ be the singlet insulating ground eigenstate; the many-body wavefunction is square-integrable over ${\mathbb R}^{3N}$ and vanishes far away from the sample. If the system is macroscopically homogeneous, the electronic term in polarization has the pretty trivial expression \[ {\bf P}^{(\rm el)} = -\frac{e}{\cal V} \me{\tilde\Psi_0}{\hat{\r}}{\tilde\Psi_0} , \qquad \hat\r = \sum_{i=1}^N \r_i  ; \label{trivial} \] the nuclear classical contribution has to be added in order to obtain a meaningful observable.

It is expedient to address the family of many-body Hamiltonians parametrized by the parameter $\kk$: \[ \hat{H}_{\kk} = \frac{1}{2m} \sum_{i=1}^N \left( \p_i + \hbar \kk \right)^2 + \hat{V}, \label{kohn} \] where $\hat{V}$ includes one-body and two-body potentials, and whose ground eigenstate is $\ket{\tilde\Psi_{0\kk}}$. In order to simplify notations we will set $\hat{H}_{0} \equiv \hat{H}$ and $\ket{\tilde\Psi_{n0}} \equiv \ket{\tilde\Psi_{n}}$. The vector $\kk$, having the dimensions of an inverse length, generalizes the Hamiltonian by including a constant vector potential: it is therefore a pure gauge. The gauge-transformed eigenstates are \[ \ket{\tilde{\Psi}_{n\kk}} = \emi{\kk \cdot \hat\r} \ket{\tilde{\Psi}_n} . \label{gauge} \] 

We pause at this point to stress an important semantical issue. The choice of the (arbitrary) $\kk$ value in \equ{kohn} fixes the gauge in the Hamiltonian. Once this fixed, there is an additional freedom in choosing the arbitrary phase factor in front of each eigenstate: even this second choice goes under the name of gauge choice. The expression of any physical observable must be gauge-invariant in both senses. I therefore alert the reader that, in the following of this paper, it is essential to realize in which context the term ``gauge'' is used.

{\color{black} By taking the $\kk$ derivative of \equ{gauge} one transforms \equ{trivial} into  \[ {\bf P}^{(\rm el)} = - \frac{i e}{\cal V} \ev{\tilde\Psi_{0\kk}| \partial_{\kk} \tilde\Psi_{0\kk}}  , \label{connek} \] 
at any $\kk$-value; in view of the subsequent developments, we set $\kk=0$ in the following:}
\[ {\bf P}^{(\rm el)} = - \frac{i e}{\cal V} \ev{\tilde\Psi_0| \partial_{\kk} \tilde\Psi_0}  . \label{conne} \]
In this seemingly innocent transformation, I have ``swept under the rug'' an issue of overwhelming importance. We start noticing that \equ{conne} has not the standard form of an observable: it is not the expectation value of an operator, at variance with \equ{trivial}. It is expressed in terms of the ground state only, as a function of $\kk$; in fact the real quantity $i \ev{\tilde\Psi_0| \partial_{\kk} \tilde\Psi_0}$ is, in the language of quantum geometry, a Berry connection evaluated at $\kk=0$. Since \equ{kohn} at two different $\kk$'s yields two different Hamiltonians, an equally acceptable gauge-transformed eigenstate would be \[ \ket{\tilde{\Psi}_{n\kk}} = \ei{\phi(\kk)} \emi{\kk \cdot \hat\r} \ket{\tilde{\Psi}_n} , \label{gauge2} \] with an arbitrary $\phi(\kk)$. 
The physical observable obtains from \equ{conne} when the gauge of \equ{gauge} is enforced; it is not allowed to adopt therein the most general gauge of \equ{gauge2}.

The gauge dependence of Berry connections is a textbook fixture of quantum geometry.\cite{Vanderbilt} In the present case \equ{conne} acquires its physical meaning only after the above specific gauge fixing. {\color{black} I stress that here is the conceptual novelty of the present work: a definition of the dipole of a bounded sample where no use is made of the position operator $\r$. The same  definition and the same gauge fixing---\equ{gauge3} below---can be exported to the PBC crystalline case.}

\section{Unbounded crystal} \label{sec:crystal}

We adopt the same Hamiltonian as in \equ{kohn}, but now within the PBC Hilbert space: the many-body wavefunction is periodic in the cubic ``supercell'' of side $L$ in each electronic variable independently, and normalized to one therein. Each Cartesian coordinate is then equivalent to the angle $\varphi_i = 2 \pi x_i / L$, and analogously for $y_i$ and $z_i$. The potential $\hat V$ enjoys the same periodicity: this means that the macroscopic field $\E$ inside the sample vanishes. We will indicate the eigenstates as $\ket{\Psi_{n\kk}}$ without a tilde, in order to distinguish them from those of the bounded crystallite; as stressed above, the multiplicative $\hat\r$ operator is ``forbidden'' in the PBC Hilbert space.\cite{rap100}

In order to address polarization, we need to ensure beforehand that the ground state is insulating. The many-body velocity operator is \[ \hat{\v}_{\kk} = \frac{1}{m} \sum_{i=1}^N (\p_i + \hbar \kk) = \frac{1}{\hbar}\partial_{\kk} \hat{H}_{\kk} ,\] hence by Hellmann-Feynman theorem the macroscopic current density is  \[ {\bf j}_{\kk} = - \frac{e}{\hbar L^3} \me{\Psi_{0\kk}}{\partial_{\kk} \hat{H}_{\kk}}{\Psi_{0\kk}} = - \frac{e}{\hbar  L^3} \partial_{\kk} {\cal E}_{0\kk} \; , \label{current} \] where ${\cal E}_{0\kk}$ is the ground-state energy. Given that an insulator does not sustain a dc current, the ground-state energy is $\kk$-independent (the opposite is true in metals).

The Hamiltonian of \equ{kohn} was first introduced in 1964 in a milestone paper by W. Kohn, who noticed that PBCs violate gauge-invariance in the conventional sense.\cite{Kohn64} {\color{black} If we try the same transformation as in \equ{gauge}, the quantity $\emi{\kk \cdot \hat\r} \ket{\Psi_0}$ is a solution of Schr\"odinger equation with energy ${\cal E}_0$, but it does not obey PBCs and therefore does not belong to the Hilbert space. }
At an arbitrary $\kk$, the genuine PBC eigenstates $\ket{\Psi_{n\kk}}$ have a nontrivial $\kk$-dependence. There is, however, a discrete set of special $\kk$ vectors for which 
    \[ {\color{black} \ket{{\Psi}_{0\kk}} = \emi{\kk \cdot \hat\r} \ket{{\Psi}_0} \label{gauge3} } \] 
obeys PBCs and yields therefore the ground eigenstate of $\hat{H}_{\kk}$: $\kk = \frac{2\pi}{L} (\ell, m, n)$, with integer $(\ell,m,n)$. 

In order to define polarization, we proceed by adopting the analogue of \equ{conne}, and in the analogous gauge. We start from the identity \[ \partial_{\kk} \mbox{ln }  \ev{\Psi_0|\Psi_{0\kk}} = \frac{\ev{\Psi_0|\partial_{\kk} \Psi_{0\kk}}}{ \ev{\Psi_0|\Psi_{0\kk}}},  \] {\color{black} which holds at any $\kk$ value;} since $\ev{\Psi_0 | \partial_{\kk} \Psi_{0}}$ is purely imaginary,
a leading-order expansion in $\kk$ yields \[ i \ev{\Psi_0 | \partial_{\kk} \Psi_{0}} \cdot \kk \simeq - \mbox{Im ln } \ev{\Psi_0|\Psi_{0\kk}} . \label{df} \] {\color{black} We pause to observe that multivaluedness debuts here. In fact \equ{df} relates two phase angles: a differential angle on the left, and a finite angle difference on the right. While a differential angle is single valued, a finite angle is defined modulo $2\pi$; upon replacing the former with the latter we are going to define a multivalued observable. We stress once more that multivaluedness is not a mathematical artifact; it is a necessary feature of polarization within PBCs.\cite{Vanderbilt}}

Next we pick a vector $\kk_1$ in the special set: $\kk_1 = \frac{2\pi}{L} (1,0,0)$, and we replace the derivative in \equ{conne} with a finite difference, in the large-sample limit:  \[  P_x^{(\rm el)} = \frac{e}{2\pi L^2} \mbox{Im ln } \ev{\Psi_0|\Psi_{0\kk_1}} . \label{rap1} \] As it stands, \equ{rap1} is gauge-dependent and cannot express an observable: it is in fact a discretized Berry connection. \equ{rap1} only acquires physical meaning when we fix the gauge by adopting the one of \eqs{gauge}{gauge3} (with no extra phase factor): \bea  P_x^{(\rm el)} &=& \frac{e}{2\pi L^2} \mbox{Im ln } \me{\Psi_0}{\emi{\kk_1 \cdot \hat\r}}{\Psi_0}  \nn &=& \frac{e}{2\pi L^2} \mbox{Im ln } \me{\Psi_0}{\emi{\frac{2\pi}{L}\sum x_i}}{\Psi_0} . \label{rap2} \eea We have thus arrived at the main message of the present work: the bounded-crystallite formula, \equ{conne}, and the crystalline formula, \equ{rap2}, are essentially the same formula, within the same gauge, in two different frameworks. 

The replacement of $\ket{\Psi_{0\kk_1}}$ in \equ{rap1} with $\emi{\kk_1 \cdot \hat\r} \ket{{\Psi}_0}$ in \equ{rap2} is allowed in insulators only. We remind that $\ket{\Psi_{0\kk}}$ obtains by following the ground state  $\ket{\Psi_{0}}$ when the $\kk$ vector in $\hat{H}_{\kk}$ is adiabatically turned on; in the metallic case---as shown by Kohn\cite{Kohn64}---the energy ${\cal E}_{0\kk}$ of such state {\it does} depend on $\kk$, and therefore  $\ket{\Psi_{0\kk_1}}$ is orthogonal to $\emi{\kk_1 \cdot \hat\r} \ket{{\Psi}_0}$. We have shown above that in the insulating case the state $\ket{\Psi_{0\kk_1}}$ has instead the same energy as $\emi{\kk_1 \cdot \hat\r} \ket{{\Psi}_0}$, and therefore  the two states may be identified.

{\color{black} The well known \equ{rap2}, sometimes dubbed ``single-point Berry phase'', was originally obtained in Ref \onlinecite{rap100} by considering a many-body Hamiltonian which is adiabatically varied in time, and showing that the time derivative of \equ{rap2} coincides with the macroscopic current density $j_x^{(\rm el)}(t)$ which flows through the insulating sample. Here I have derived the same result via a different logic: polarization itself obtains without addressing currents at all, starting instead from an unconventional definition of the dipole of a bounded sample. } 

Finally, the nuclear term in polarization can be added to \equ{rap2} in a very compact form. If the nuclei of charge $Z_\ell$ sit at sites $\R_\ell$ in the supercell, the expression is \[  P_x = \frac{e}{2\pi L^2} \mbox{Im ln } \me{\Psi_0}{\ei{\frac{2\pi}{L}(\sum_\ell Z_\ell X_\ell -\sum_i x_i)}}{\Psi_0} , \label{rap3} \] where $X_\ell = R_{\ell,x}$. Owing to charge neutrality, polarization is invariant by translation of the coordinate origin (as it must be). It is argued that \equ{rap3} also holds when the quantum nature of the nuclei is considered.

\section{Multivalued polarization in crystals} \label{sec:multi}

Bulk polarization is a lattice, not a vector,  and in fact the main entry of \eqs{rap2}{rap3} is the multivalued function ``Im ln''. But it is also clear that for a three-dimensional system these equations cannot be accepted as they stand in the large-sample limit: the prefactor goes in fact to zero. {\color{black} It has been shown in Ref. \onlinecite{Ortiz94} that, by exploiting crystalline symmetry, \eqs{rap2}{rap3} eventually yield an uniquely defined multivalued observable; I take the present occasion for providing a somewhat more intuitive proof.} 

By definition, whenever a material is crystalline, a uniquely defined lattice can be associated with the real sample. The lattice is a ``mathematical construction'',\cite{Kittel} uniquely defined---by means of an appropriate average---even in cases with correlation, finite temperature, quantum nuclei, chemical disorder (i.e. crystalline alloys, a.k.a. solid solutions), where the actual wavefunction may require a supercell (multiple of the primitive lattice cell). 

We consider---without loss of generality---a simple cubic lattice of constant $a$, where the supercell side $L$ is an integer multiple of $a$: $L=Ma$. Suppose the potential $\hat V$ in the Hamiltonian is adiabatically varied in time; we define the phase angle \[ \gamma_x(t)  =  
\mbox{Im ln } \me{\Psi_0(t)}{\ei{\frac{2\pi}{L}(\sum_\ell Z_\ell X_\ell -\sum_i x_i)}}{\Psi_0(t)} , \] where $\ket{\Psi_0(t)}$ is the adiabatic ground eigenstate. The current flowing across a section of area $L^2$ normal to $x$ is \[ I_x(t) = L^2 \dot P_x(t) = \frac{e}{2\pi} \dot\gamma_x(t) .\] Owing to cristalline periodicity, The current $I_x(t)$ is the sum of $M^2$ identical currents, each flowing through a microscopic section of area $a^2$; one can therefore define a reduced crystalline phase angle
$\gamma_x^{(\rm crystal)}$ such that $ \dot\gamma_x(t) = M^2 \dot\gamma_x^{(\rm crystal)}(t)$. The crystalline polarization is thus expressed in terms of $\gamma_x^{(\rm crystal)}$ as \[ P_x = \frac{e}{2\pi a^2} \gamma_x^{(\rm crystal)} ; \] the case of independent electrons is presented in detail in the next Section.

A generic lattice is dealt with by means of a coordinate transformation;\cite{rap_a12} the bulk value of ${\bf P}$ is then ambiguous modulo $e\R/{\cal V}_{\rm cell}$, where $\R$ is a lattice vector and ${\cal V}_{\rm cell}$ is the volume of a primitive cell. The quantity $e\R/{\cal V}_{\rm cell}$ goes under the name of polarization ``quantum''.
By definition a primitive cell is a minimum-volume one:\cite{Kittel} this choice is mandatory in order to make ${\bf P}$ an unambiguously defined multivalued observable. Finally we observe that the modulo ambiguity is only removed when the termination of the bounded sample is specified; it is also required that even the surfaces, as well as the bulk, are insulating.\cite{Vanderbilt} Insofar as the crystalline system is unbounded the modulo ambiguity cannot be removed.

\section{Single-determinant wavefunction} \label{sec:single}

Within mean field (either Hartree-Fock or Kohn-Sham) the ground eigenstate $\ket{\Psi_0}$  in the Schr\"odinger representation is a Slater determinant of $N/2$ doubly occupied orbitals; in the crystalline case translational symmetry allows choosing the orbitals in the Bloch form. For the sake of simplicity we get rid of trivial factors of two, by considering a Slater determinant of singly occupied orbitals (so-called ``spinless electrons''); furthermore we consider the contribution to $P_x^{(\rm el)}$ of a single occupied band. 

In the simple cubic case, as dealt with above, the Bloch vectors are: \[ \k_m = \frac{2\pi}{Ma}(m_1,m_2,m_3), \quad m_s = 0,1,\dots,M-1, \] 
where $m \equiv (m_1,m_2,m_3)$.
The Bloch orbitals $\ket{\psi_{\k_m}} = \ei{\k_m \cdot \r} \ket{u_{\k_m}}$ are normalized over the crystal cell of volume $a^3$. It is expedient to define the auxiliary Bloch orbitals $\ket{\phi_{\k_m}} = \ei{\frac{2\pi}{L}\, x } \ket{\psi_{\k_m}}$, and $\ket{\Phi_0}$ as their Slater determinant; we also define ${\bf q}= (\frac{2\pi}{Ma},0,0)$. Then \[ \me{\Psi_0}{\ei{ \sum_i{\bf q} \cdot  \r_i }}{\Psi_0}   = \ev{\Psi_0|\Phi_0} = \frac{1}{M^{3N}}\mbox{det } {\cal S} , \] where ${\cal S}$ is the $N \times N$ overlap matrix of the orbitals, in a different normalization: \bea {\cal S}_{mm'} &=& M^3 \ev{\psi_{\k_m}|\phi_{\k_{m'}}} = M^3\me{u_{\k_m}}{\ei{({\bf q} + \k_{m'} - \k_m) \cdot \r}}{u_{\k_{m'}}} 
 \\ &=&  M^3 \ev{u_{\k_m} | u_{\k_{m'}}} \, \delta_{{\bf q} + \k_{m'} - \k_m} = M^3 \ev{u_{\k_m}|u_{\k_{m}-{\bf q}}}  \delta_{mm'}. \nonumber
\eea The normalization factors cancel: we have in fact \[ \me{\Psi_0}{\ei{\frac{2\pi}{L}\, \sum_i x_i }}{\Psi_0}  = \frac{1}{M^{3N}}\mbox{det } {\cal S} = \prod_{m_1,m_2,m_3 =0}^{M-1} \ev{u_{\k_m}|u_{\k_{m}-{\bf q}}} , \] \bea \gamma_x^{(\rm crystal)} &=& \frac{1}{M^2} 
\mbox{Im ln } \me{\Psi_0}{\emi{\frac{2\pi}{L}\, \sum_i x_i }}{\Psi_0} \nn  &=& -\frac{1}{M^2} \sum_{m_2,m_3 =0}^{M-1} \mbox{Im ln } \prod_{m_1 =0}^{M-1} \ev{u_{\k_m}|u_{\k_{m}-{\bf q}}} . \eea This is indeed the single-band version of the discretized Berry-phase formula routinely implemented in ab-initio electronic-stucture codes for computing macroscopic polarization;\cite{Vanderbilt} the classical nuclear term has to be added.

\section{Conclusions} \label{sec:conclusions}

The theory of polarization in condensed matter was developed along the 1990s\cite{rap73,King93,Vanderbilt93,Ortiz94,rap100} and is now a staple of electronic structure theory. The relevant formulas adopt concepts from quantum geometry,\cite{Vanderbilt} and have no relationship to the (trivial) formula for the dipole of a bounded sample; this owes to the fact that the multiplicative position operator $\r$ is no longer a legitimate quantum-mechanical operator when the boundary conditions of condensed matter physics are adopted.\cite{rap100}

Here I show that it is possible to alternatively express the dipole of a bounded sample, making no use of the $\r$ operator: such expression has the virtue of being adoptable almost as such even in the case of an unbounded crystalline sample. The resulting formula defines bulk crystalline polarization as a multivalued observable. I also show that, in the special case of a single-determinant many-body wavefunction (either Hartree-Fock or Konhn-Sham), one retrieves the algorithm currently implemented in several electronic-structure codes.\cite{Vanderbilt}

Finally, {\color{black} it is worth observing} that---when the large-crystallite limit is ideally taken---one obtains the value of crystalline polarization only if the limit is taken in the appropriate way. The reason is that the crystalline formula assumes by construction zero macroscopic field, while instead a nonzero macroscopic field (depolarization field) is in general present inside a polarized---either pyroelectric or ferroelectric---macroscopic sample in vacuo.

\medskip

\section*{Acknowledgments}
Work supported by the ONR Grant No. N00014-17-1-2803.

Data sharing is not applicable to this article as no new data were created or analyzed in this study.


\begin{thebibliography}{10}

\bibitem{Kittel}
{ C. Kittel, {\it Introduction to Solid State Physics}, 8th. edition (Wiley,
  Hoboken, NJ, 2005)}.

\bibitem{rap73}
{ R. Resta, Ferroelectrics {\bf 136}, 51 (1992)}.

\bibitem{King93}
{ R. D. King-Smith and D. Vanderbilt, Phys. Rev. B {\bf 47}, 1651 (1993)}.

\bibitem{Vanderbilt93}
{ D. Vanderbilt and R. D. King-Smith, Phys. Rev. B {\bf 48}, 4442 (1993)}.

\bibitem{Ortiz94}
{ G. Ort\'{\i}z and R. M. Martin, Phys. Rev. B {\bf 49}, 14202 (1994)}.

\bibitem{rap100}
{ R. Resta, Phys. Rev. Lett. {\bf 80}, 1800 (1998)}.

\bibitem{Spaldin12}
{ N. A. Spaldin, J. Solid State Chem. {\bf 195}, 2 (2012)}.

\bibitem{Vanderbilt}
{ D. Vanderbilt, {\it Berry Phases in Electronic Structure Theory} (Cambridge
  University Press, Cambridge, 2018)}.

\bibitem{rap136}
{ K. N. Kudin, R. Car, and R. Resta, J. Chem. Phys. {\bf 127}, 194902 (2007)}.

\bibitem{Kohn64}
{ W. Kohn, Phys. Rev. {\bf 133}, {A171} (1964)}.

\bibitem{rap155}
{ R. Resta, Eur. Phys J. B {\bf 91}, 100 (2018)}.

\bibitem{rap_a12}
{ R. Resta, Rev. Mod. Phys. {\bf 66}, 899 (1994)}.

\end{thebibliography}

\end{document}